\documentstyle[12pt]{article}
\def\ea{{\it et al.} }
\def\ref{\par\noindent\hangindent 20 pt}

\begin{document} 

\input{psfig.tex}

\begin{center}
{\bf On the Relation Between Radio and Non-Radio Elliptical Galaxies}

{\bf Riccardo Scarpa \& C. Megan Urry}

Space Telescope Science Institute 
\end{center} 

\begin{center}
{\bf Abstract}
\end{center}

Empirical evidence suggests that elliptical galaxies hosting a
radio source may not be different from normal non-radio ellipticals.  To
test this possibility, we use Monte Carlo simulations to reproduce the
distribution of radio galaxies in the radio-optical luminosity plane.
The input parameters of the simulation are the optical luminosity
function (LF) of ellipticals and the radio LF of radio galaxies, linked
by a function giving the probability for an elliptical to be a radio
galaxy.

Simulations reproduce the observations well, supporting unification of
radio and non-radio ellipticals, provided that the probability
of an elliptical hosting a radio source is proportional to
the square of its optical luminosity. 

The difference of $\sim 0.5$ mag in
average optical luminosity between FRI and FRII radio galaxies is also
explained in this framework.

\section{Introduction}

Regardless of nuclear activity, all ellipticals
lie in the same fundamental plane, have similar ellipticity
distributions, isophotal twists, and colors (Homabe \& Kormendy 1987;
Ledlow \& Owen 1995; Govoni \ea 2000; Urry \ea 2000).
Furthermore,
the presence of massive black holes at the centers of
elliptical galaxies is the rule rather than the exception (Ho 1998;
Magorrian \ea 1998; Richstone \ea 1998; van der Marel 1999). 
This suggests all ellipticals have the potential of
experiencing a phase of intense nuclear activity.
 
To test this attractive possibility, we used 
Monte Carlo simulations to check whether
observed samples of radio galaxies can be random selections of elliptical
galaxies.

 We start from the optical luminosity function (LF) of ellipticals, which 
in this scenario gives the number of potential radio galaxies 
as a function of optical magnitude. Then we introduce a
probability function for the fraction of galaxy, of a given
magnitude, to host a radio source.

\section{Calculation}

Based on available empirical results for radio galaxies,
the following general assumptions are made:

\begin{enumerate}

\item
The distribution of ellipticals in
optical luminosity $L$ is given by a Schechter function,
with $M_R^*=-22.8$ mags and $\alpha=+0.2$, as found 
in the Stromlo-APM experiment (Loveday \ea 1992).

\item
All elliptical galaxies of all optical luminosities 
have the potential of being radio sources, with probability 
$S(L)= S^* (\frac{L}{L^*})^h$, where $S^*$ sets the overall normalization.
From the bivariate LF it is known that $S(L)\propto L^2$ 
(Ledlow \& Owen 1996), so we set $h=2$ (the results are not very 
sensitive to the exact value of $h$).

\item
Regardless of their optical luminosity, 
all active ellipticals produce radio sources with total power $P$
distributed following the known radio luminosity function
$\frac{dN}{dP} \propto P^{\beta}$, with $\beta = -2$ 
(Auriemma \ea 1977; Toffolatti \ea 1987; Urry \& Padovani 1995; Ledlow
\& Owen 1996).

\item
In the radio-optical luminosity plane, FR I and FR II are separated by
a transition line roughly proportional to $L^2$ (Bicknell 1995).

\end{enumerate}

The number of radio sources per unit volume, having
optical luminosity $L$ is the product
of the optical LF times the probability $S$ (points 1 and 2):

\begin{equation}
N(L)= \Phi(L) S(L) = \frac{\Phi ^*}{L^*} S^* (\frac{L}{L^*})^{(\alpha+h)}
e^{(\frac{L}{L^*})}.
\end{equation}

\noindent
This is a Schechter function with exponent $(\alpha + h)=2.2$. 
The final distribution of radio sources in the radio-optical 
luminosity plane, as derived in a radio-flux-limited survey,
is given by the product of $N(L)$ times the function 
$\frac{dN}{dP}$, assumed to be the same for all optical luminosities (point 3),
times the volume $V(P)$ over which sources of power $P$ can be observed
above the flux limit of the survey.
Once a random set of galaxies has been generated, sources
are divided into FR I and FR II according to point 4.

Given the assumptions, there are essentially no free parameters, but we note 
the exponent $h$ is not well determined. Also,
different groups have found
significantly different values of $M^*$ and $\alpha$ (Muriel \ea 1995;
Lin \ea 1996, Loveday \ea 1992; Zucca \ea 1997), and there is some
freedom in the position of the FR I--II transition power.

\begin{figure}
\centerline{
\psfig{file=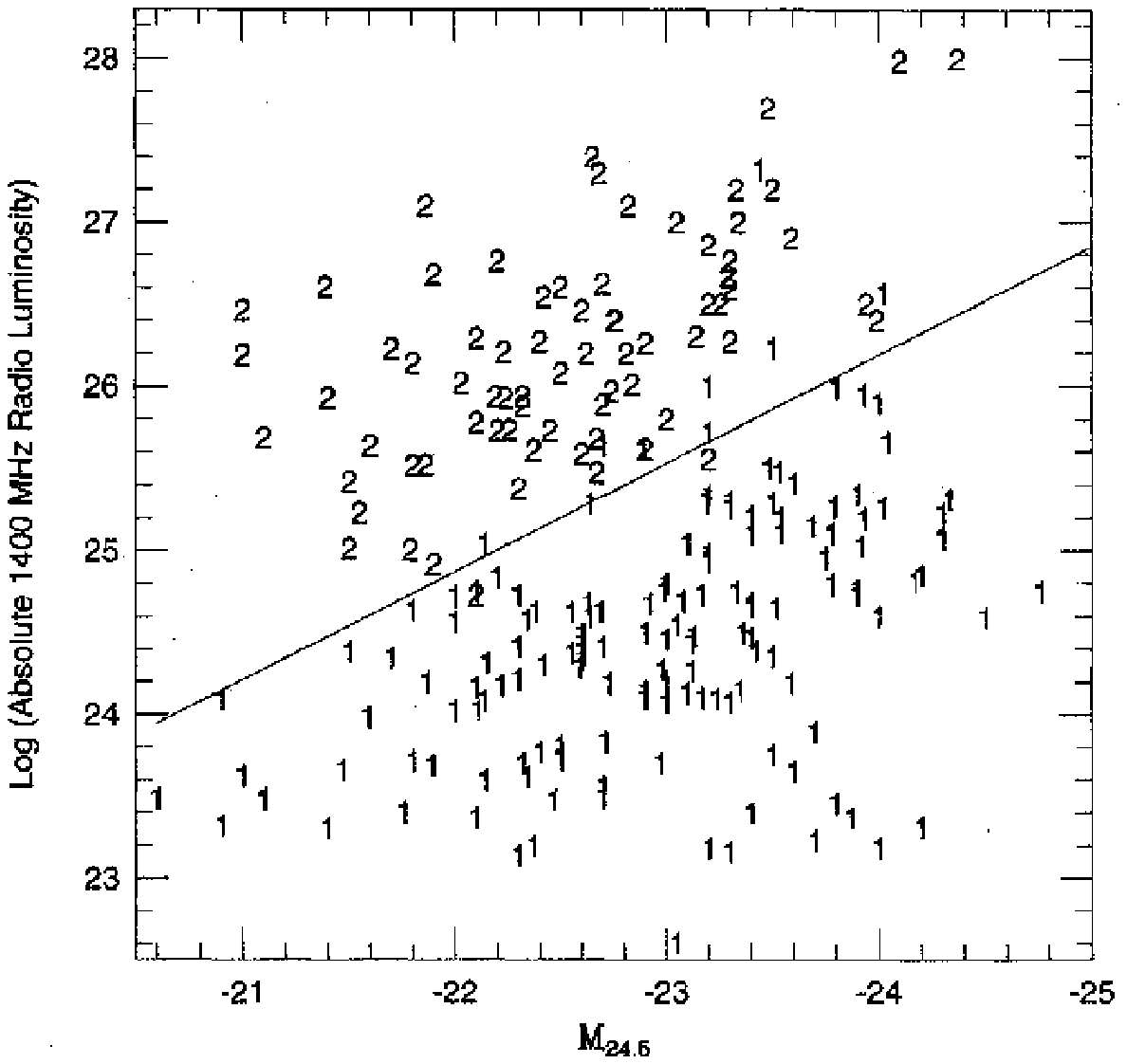,width=0.5\linewidth}
\psfig{file=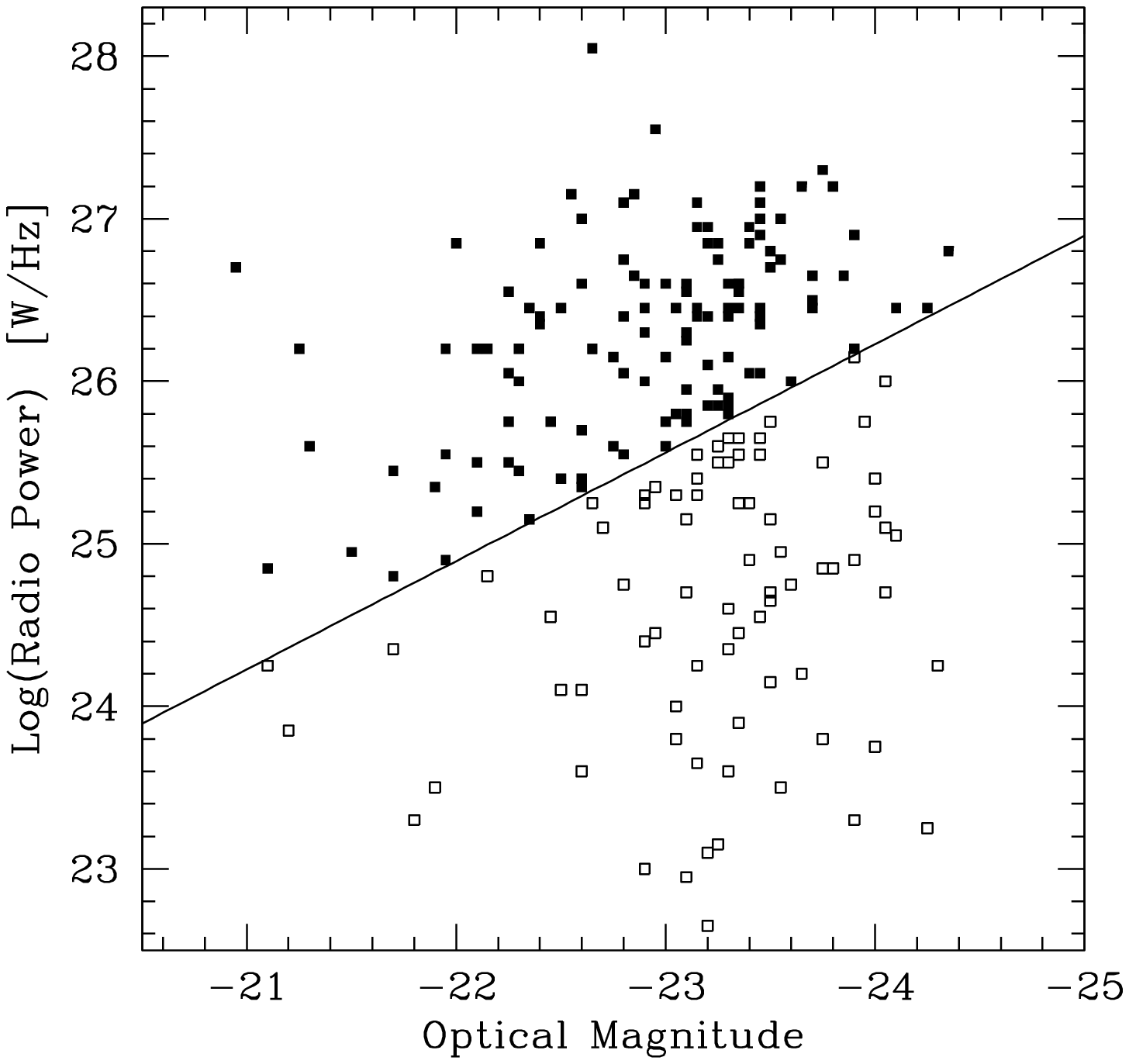,width=0.485\linewidth}
}
\caption{
{\bf Comparison of our simulation with data from Figure 1 of Ledlow \& Owen (1996).}
(For consistency, this figure is computed with $H_0=75$ km/s/Mpc.)
{\bf Right Panel:} The distribution of radio galaxies (1=FRI and 2=FRII)
in the radio-optical
luminosity plane, as derived by Ledlow \& Owen (1996), plotting 
data  extracted from a complete flux-limited survey, 
to 0.1 Jy at 1.4 GHz, with no redshift limit. The solid line separating
FR I from FR II is the one originally used by Ledlow \& Owen.
{\bf Left Panel:} Representative Monte Carlo simulation matched to 
the Ledlow \& Owen (1996) sample. The simulation
nicely reproduces the almost
uniform coverage of the plane in the region $-25<M_R<-21$ mag and
$23<Log(P)<28$ W/Hz, with maximum concentration around the center of
this region. Solid squares represent FR II, open squares FR I.
The observed distribution occurs because:
(1)
brighter galaxies are not present because of the upper cuttoff in
the optical LF;
(2)
no galaxies are observed fainter than $M_R\sim -21$ mag  because the function
N(L) decreases rapidly at low optical luminosities;
(3)
at low radio power, the number of galaxies observed is limited by the radio flux 
cutoff of the  
survey and by the small volume over which these faint objects can be discovered.
(4)
and at high radio power the limit is set by the rapidly 
decreasing probability of having radio sources of high powers.
}
\end{figure}
\begin{figure}
\centerline{
\psfig{file=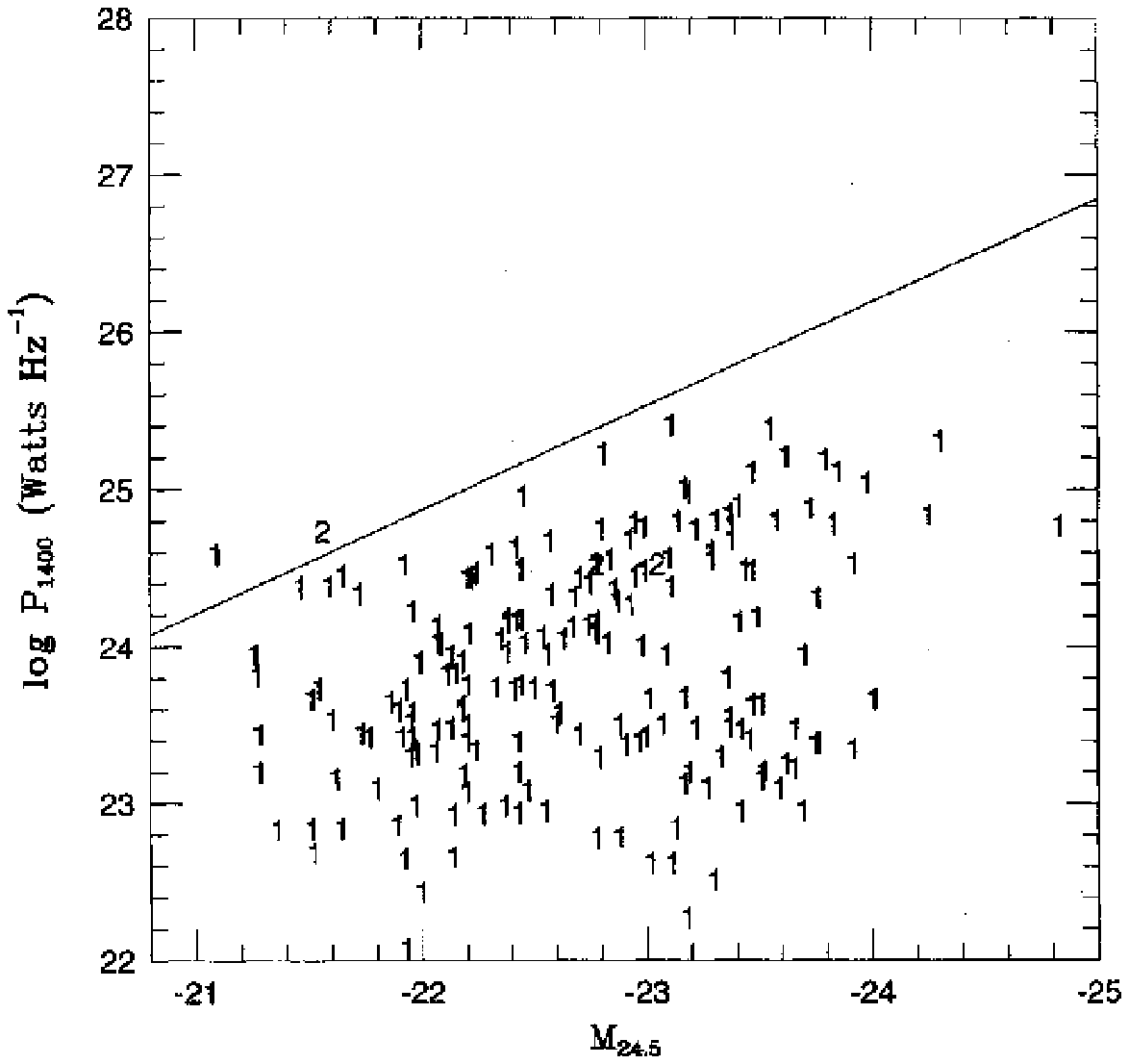,width=0.48\linewidth}
\psfig{file=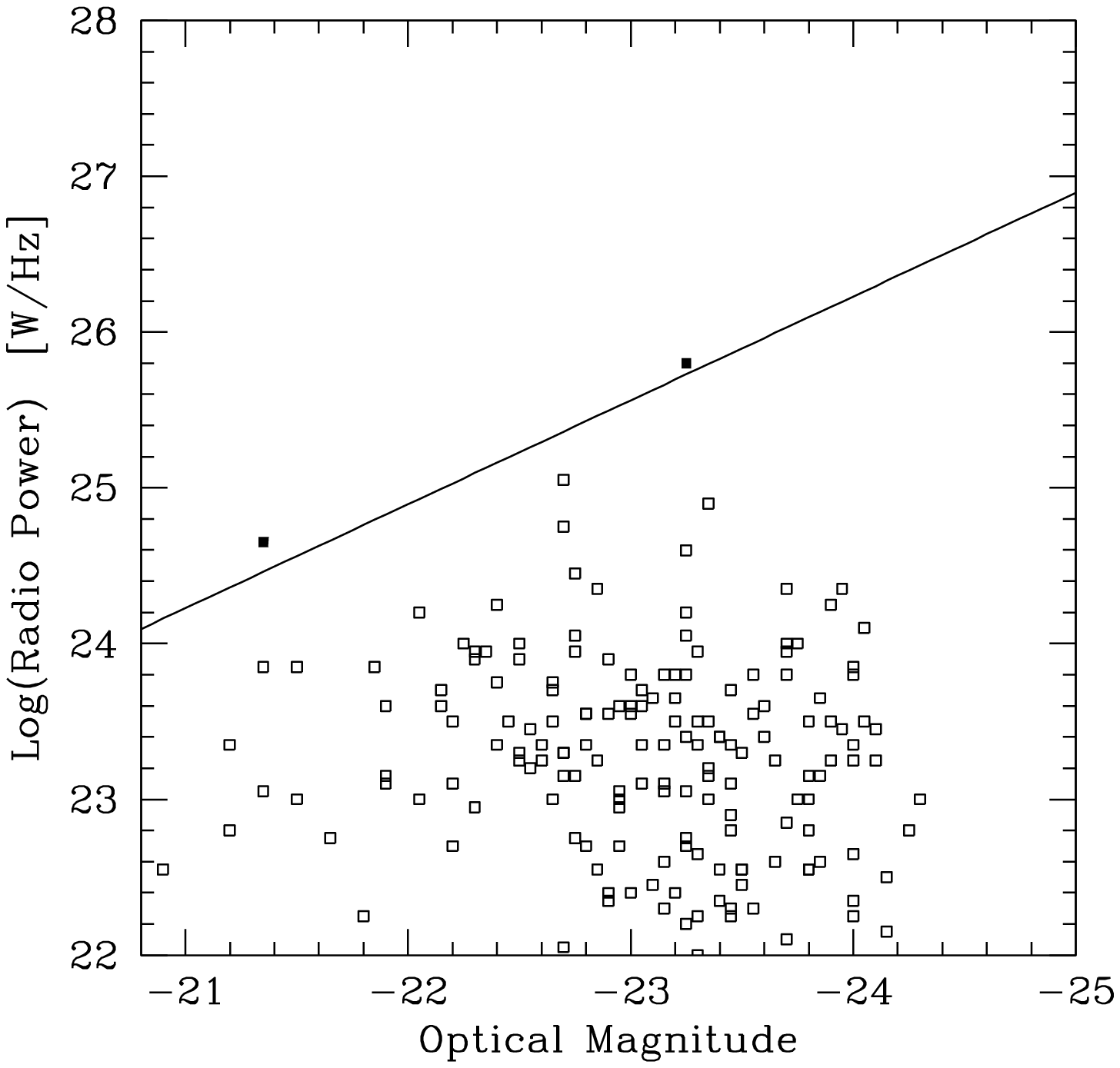,width=0.48\linewidth}
}
\caption{
{\bf Comparison with Figure 3 from Ledlow \& Owen (1996).}
(For consistency, this figure is computed with $H_0=75$ km/s/Mpc.)
{\bf Left Panel:} Radio and optical luminosity for a
sample of 188 radio sources, complete out to $z=0.09$ down to a radio
flux of 0.01 Jy at 1.4 GHz, from Ledlow \& Owen (1996).  
{\bf Right Panel:} Distribution derived from our Monte Carlo simulations.
As in the real data,
there are no very bright radio sources because of the small volume
surveyed. In particular, basically all sources are below the
transition line and should be FR I (open squares), as indeed 
observed by Ledlow \& Owen. 
}
\end{figure}

\begin{figure}
\centerline{
\psfig{file=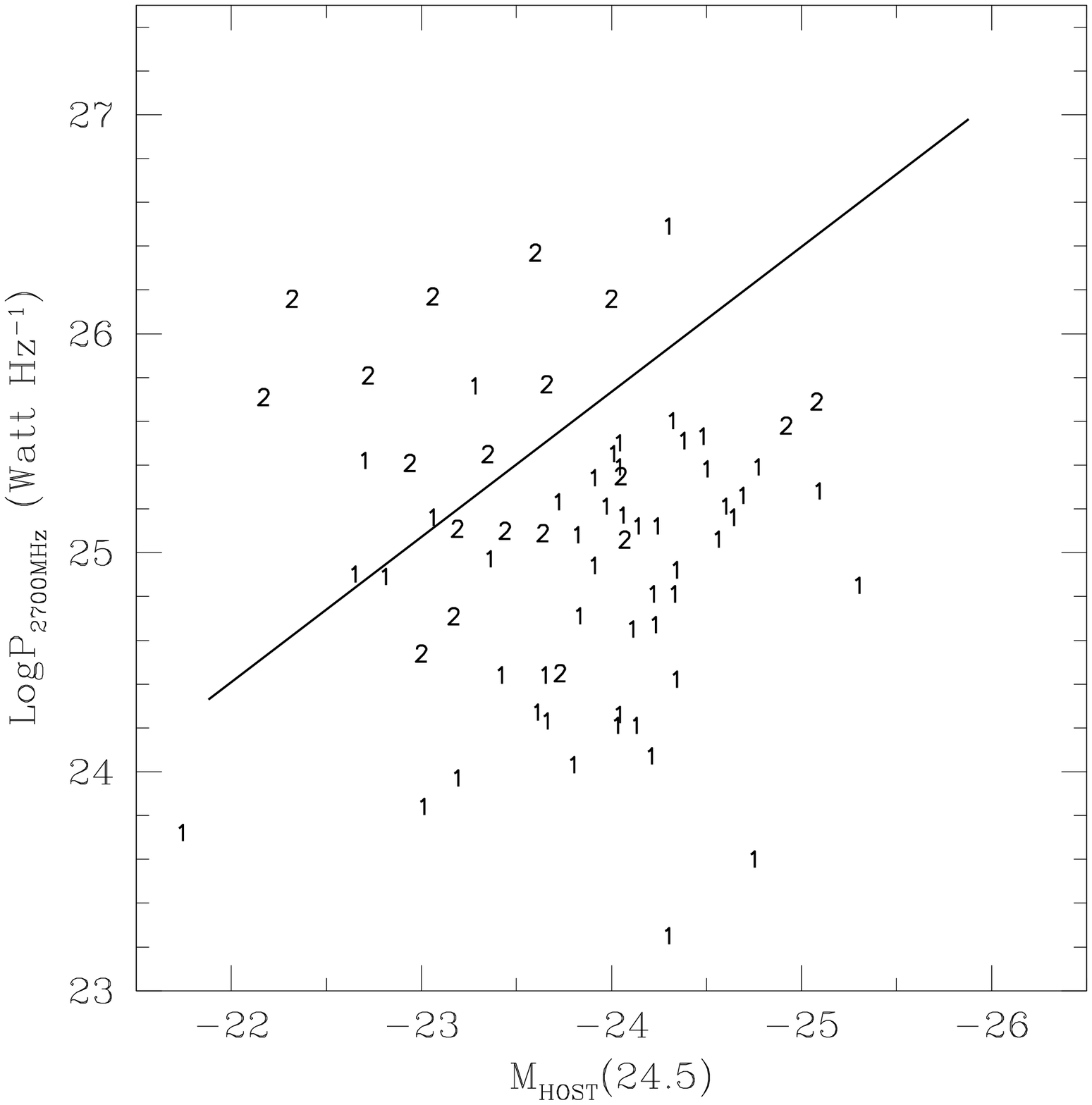,width=0.45\linewidth}
\psfig{file=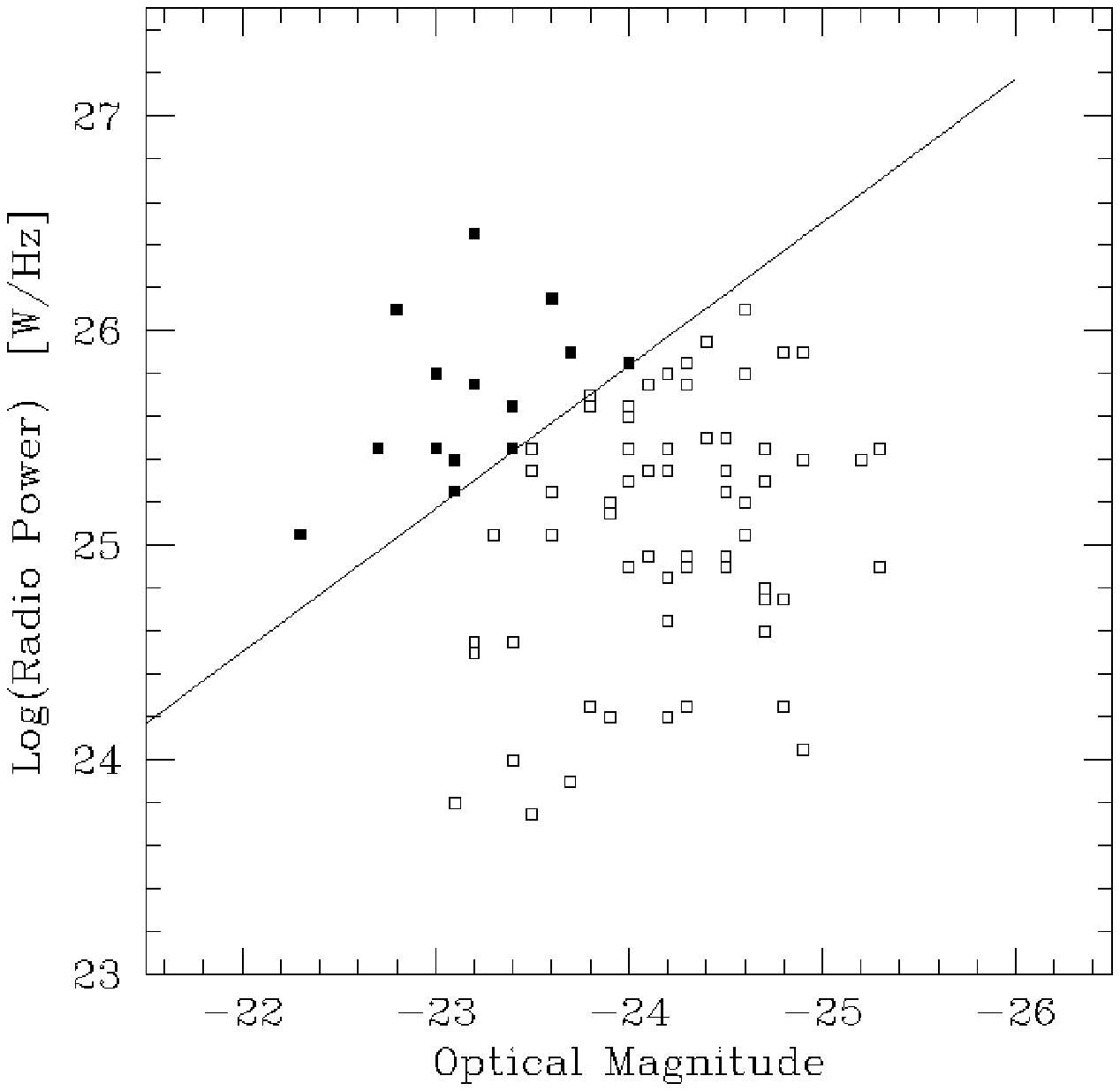,width=0.46\linewidth}
}
\caption{
{\bf Comparison with data from Govoni et. al. (2000).}
(For consistency, this figure is computed with $H_0=50$ km/s/Mpc.)
{\bf Left Panel:} Distribution of radio galaxies studied by Fasano,
Falomo \& Scarpa (1996), with final results presented by Govoni \ea
(2000). The original sample includes all radio galaxies in the redshift range
$0.01<z<0.12$, down to a flux limit, at 2.7 GHz,  of 2 Jy 
for part of the sample and 0.25 Jy for the rest.
{\bf Right Panel:}
Result of a Monte Carlo simulation matched to the Govoni \ea sample.
The agreement is excellent. We are able
to explain nicely the distribution of the sources in the radio-optical
luminosity plane, as well as the relative population of FR I (open squares)
and FR II (solid squares).  
}
\end{figure}

\begin{figure}
\psfig{file=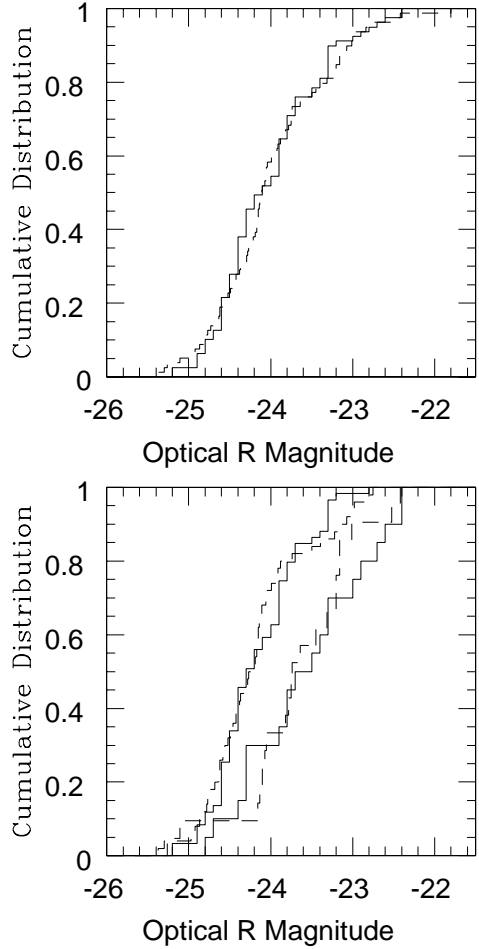,width=7cm}
\caption{
Distributions of optical magnitudes from Govoni \ea (2000)
{\bf Left Panel:} Cumulative distribution of absolute magnitudes for
the real (dashed line) and simulated (solid line) data set shown in Fig. 3. 
{\bf Right Panel:} Cumulative distribution of absolute magnitudes for
FR I (Left) and FR II (Right) radio galaxies separately.  
The agreement is excellent, explaining the difference in average optical
luminosity between FR I and FR II as a subtle selection effect.
}
\end{figure}

\section{Results and Discussion}

Result shows that under quite
general assumption the observed distribution of radio galaxies in the
radio-optical luminosity plane is nicely reproduced, as is the
observed difference in optical luminosity between FR I and FR II.

The physical basis for our result is that all
ellipticals should have a central black hole 
(van der Marel 1999; Macchetto \ea 1999).
Active and non-active galaxies are linked by a
probability function, found to be $\propto L^2$.  We do not
attempt to explain why radio sources are preferentially observed in
giant ellipticals, however, the $L^2$ dependence 
of the probability of radio activity comes from the observed shape of the
bivariate radio LF, and is similar to the dependence of the
transistion power from FR I to FR II. There must be a deeper physical
meaning for this.

Once accretion onto the black hole has began, the strength of the 
radio emission should depend mostly on the accretion rate, 
which is independent from the galaxy size, justifying
assumption 3.

The $\sim L^2$ dependence of the transition power from FR I to FR II 
imposes that to be an FR II, a bright galaxy must be associated with
a very powerful radio source, a very improbable combination given the
steepness of both radio and
optical LF. Thus, from probability alone, the association FR II -- faint 
galaxies is favored,
causing the difference in observed optical
luminosity between the two classes. No deeper
physical difference between FR I and FR II host galaxies is required.

\bigskip

\noindent
{\bf References}

 \noindent
Auriemma C., Perola G.C., Ekers R., \ea 1977, A\&A 57, 41

 \noindent
Bicknell G.V. 1995, ApJS 101, 29

 \noindent
Fasano G., Falomo R. \& Scarpa R. 1996, MNRAS 282, 40

 \noindent
Govoni F., Falomo R., Fasano G. \& Scarpa R. 2000, A\&A 353, 507

 \noindent
Ho L.C. 1998, in ``Observational Evidence for Black Holes in\\ 
\indent
the Universe'', S.K. Chakraberti, ed. Kluwer, p.157

 \noindent
Homabe M. \& Kormendy J. 1987, in ``Structure and Dynamics of Galaxies'',\\
\indent
IAU symp. N. 127, p. 379, ed. de Zeeuw, Reidel, Dordrecht. 

 \noindent
Ledlow M.J. \& Owen F.N. 1995, AJ 109, 853

 \noindent
Ledlow M.J. \& Owen F.N. 1996, AJ 112, 9

 \noindent
Lin H., Kirshner R.P., Shectman S.A. \ea 1996, ApJ 464, 60

 \noindent
Loveday J., Peterson B.A., Efstathious G. \& \\
\indent
Maddox S.J. 1992, ApJ 390, 338

 \noindent
Macchetto F.D. 1999, in press (Astro-ph 9910089) 

 \noindent
Magorrian J., Tremaine S., Richstone D. \ea 1998, AJ 115, 2285

 \noindent
Muriel H., Nicotra M.A. \& Lambas D.G. 1995, AJ 110, 1032

 \noindent
Richstone D., Ajhar E.A., Bender R. \ea 1998, Nature 395, 14

 \noindent
Toffolatti L., Franceschini A., Danese L \& de Zotti G. 1987, A\&A 184, 7

 \noindent
Urry C.M. \& Padovani P. 1995, PASP 107, 803

 \noindent
van der Marel R. 1999, AJ 117, 744

 \noindent
Zucca E., Zamorani G., Vettolani G., \ea 1997, A\&A 326. 477

\end{document}